\documentclass[showpacs,preprintnumbers,amsmath,amssymb,aps,twocolumn,nofootinbib]{revtex4-1}

\usepackage{graphicx}% Include figure files
\usepackage{dcolumn}% Align table columns on decimal point
\usepackage{color}

\usepackage{hyperref}
\hypersetup{%
colorlinks=true,%
linkcolor=blue,
citecolor=blue,
urlcolor=magenta
}
\newcommand{\sumint}{\mbox{$\sum$}\kern-2.7ex\int}

\def\gtsim{\mathrel{\hbox{\raise0.2ex
\hbox{$>$}\kern-0.75em\raise-0.9ex\hbox{$\sim$}}}}
\def\ltsim{\mathrel{\hbox{\raise0.2ex
\hbox{$<$}\kern-0.75em\raise-0.9ex\hbox{$\sim$}}}}

\begin{document}

%\preprint{xxxx}

\title{A renormalization group improvement for thermally resummed effective potential}

\author{Koichi Funakubo$^{1}$}
\email{funakubo@cc.saga-u.ac.jp}
\author{Eibun Senaha$^{2,3}$}
\email{eibunsenaha@vlu.edu.vn (corresponding author)}
\affiliation{$^1$Department of Physics, Saga University,
Saga 840-8502 Japan}
\affiliation{$^2$Subatomic Physics Research Group, Science and Technology Advanced Institute, Van Lang University, Ho Chi Minh City, Vietnam}
\affiliation{$^3$Faculty of Applied Technology, School of Technology, Van Lang University, Ho Chi Minh City, Vietnam}

\date{\today}

\begin{abstract}
We propose a novel method for renormalization group improvement of thermally resummed effective potential.
In our method, $\beta$-functions are temperature dependent as a consequence of the divergence structure in resummed perturbation theory. In contrast to the ordinary $\overline{\text{MS}}$ scheme, the renormalization group invariance of the resummed finite-temperature effective potential holds order by order, which significantly mitigates a notorious renormalization scale dependence of phase transition quantities such as a critical temperature even at the one-loop order. We also devise a tractable method that enables one to incorporate temperature-dependent higher-order corrections by fully exploiting the renormalization group invariance.   
\end{abstract}

%\pacs{05.70.Fh, 11.10.Wx, 12.38.Cy, 98.80.Cq}
%11.10.Wx 	Finite-temperature field theory 
%12.38.Cy 	Summation of perturbation theory 
%05.70.Fh 	Phase transitions: general studies
%98.80.Cq 	Particle-theory and field-theory models of the early Universe

\maketitle

%\tableofcontents

%---------------------------------------------------------- Section 1 ----------------------------------------------------------
\section{Introduction}
%----------------------------------------------------------------------------------------------------------------------------------
Thermal effective potential has been widely used to analyze phase transitions such as electroweak phase transition (EWPT). As is well known, the perturbative method to evaluate the effective potential suffers from bad high-temperature behavior even in a theory with small coupling constants~\cite{Dolan:1973qd,Linde:1980ts}. One of the remedies is to incorporate the most dominant part of the higher-order terms at high temperatures, that is, the mass corrections which are proportional to $T^2$, into the lower-order contributions in a systematic manner. This is the so-called resummation of the thermal mass, or, simply, the thermal resummation. The resummation also cures the infrared divergence originating from the zero Matsubara frequency mode in bosonic-loop contributions.\footnote{Since the smallest frequency of a fermion is $\pi T$, the effect of the fermionic thermal resummation is much weaker than the bosonic one, so one usually considers only the bosonic thermal resummation.}
The renormalization-group (RG) improvement of the effective potential is another method to rearrange the perturbation series, in which some of higher-oder contributions are taken into the lower-order terms in perturbation theory~\cite{Coleman:1973jx,Kastening:1991gv,Bando:1992np,*Bando:1992wy,Ford:1992mv}. It is based on the fact that the bare Lagrangian, hence, the all-order results including the counterterms (CTs), be independent of the renormalization scale. Although the perturbative effective potential has an explicit scale dependence at some fixed order, the scale invariance is improved by introducing the running parameters. Once the effective potential is made scale invariant at some order, the scale can be fixed in such a way that some of the higher-order terms vanish. The running parameters defined by use of $\beta$-functions have been often determined by renormalizing the theory with the $\overline{\text{MS}}$-scheme. At finite temperatures, a new scale-dependent term arises, which cannot be taken care of by the running parameters defined by the $\overline{\text{MS}}$-scheme. This situation is made more serious, when one executes the thermal resummation, leading to violation of the order-by-order RG invariance of the effective potential (for other approaches, see, e.g., Refs.~\cite{Gould:2021oba,*Croon:2020cgk,*Schicho:2022wty,Athron:2022jyi}).

In this letter, we propose a novel RG improvement method for the resummed effective potentials in which the RG invariance holds order by order. In our method, $\beta$-functions are properly defined in resummed perturbation theory instead of using those in the $\overline{\text{MS}}$ scheme. As a consequence, our $\beta$-functions of the dimensionful parameters are temperature dependent. For illustrative purpose, we first work in the $\phi^4$ theory and explicitly show the RG invariance of the resummed one- and two-loop effective potentials in our scheme. Moreover, we further refine the effective potential by incorporating a series of dominant temperature-dependent higher-order terms by taking advantage of the RG invariance. 
To apply our scheme to a case of first-order phase transition as needed for electroweak baryogenesis~\cite{Kuzmin:1985mm} (for reviews see, e.g., Refs.~\cite{Rubakov:1996vz,*Funakubo:1996dw,*Riotto:1998bt,*Trodden:1998ym,*Bernreuther:2002uj,*Cline:2006ts,*Morrissey:2012db,*Konstandin:2013caa,*Senaha:2020mop}), we extend the $\phi^4$ theory by adding another real scalar field. We make numerical comparisons between the $\overline{\text{MS}}$ and our schemes and show that the latter yields much less renormalization scale dependence on a critical temperature even at the one-loop level. At the two-loop level, however, not much numerical differences are observed between the two schemes unless hard thermal loops are significantly sizable. Our numerical study also shows that our refined RG-improved one-loop effective potential can capture the two-loop order effects properly.
This would be particularly useful when the two-loop effective potential is not available. 

%---------------------------------------------------------- Section 2 ----------------------------------------------------------
\section{$\beta$-functions in the resummed theory}\label{sec:beta}
%----------------------------------------------------------------------------------------------------------------------------------
We first clarify differences between $\beta$-functions in the $\overline{\text{MS}}$ and those in our scheme. To make our discussion simpler, we focus on scalar theories. The derivation of $\beta$-functions in more general theories is given in Ref.~\cite{Funakubo:2023eic} (see also Ref.~\cite{Machacek:1983tz,*Machacek:1983fi,*Machacek:1984zw}). Let us collectively denote scalar fields and couplings as $\phi_i(x)$ and $g_k$ and scalar masses as $m_a^2$. We also define a vacuum energy as $\Omega$. As we see in the next section, $\Omega$ is also needed to show the RG invariance of the effective potential.

In this work, we adopt the dimensional regularization in which the spacetime dimension is analytically continued to $d=4-\epsilon$ dimension. Since the $\beta$-functions of the dimensionless parameters are not affected within the scope of our discussion here, we derive only those of dimensionful parameters. In the $\overline{\text{MS}}$ scheme, the bare parameters are expressed in terms of the renormalized ones and $\epsilon$ poles as
\begin{align}
m_{Ba}^2 & = \left(\delta_{ab}+\sum_{n=1}^\infty\frac{b_{ab}^{(n)}(g)}{\epsilon^n} \right)m_b^2,\label{mB} \\
\Omega_B\mu^\epsilon  & = \Omega+\sum_{n=1}^\infty\frac{\omega_n(g)}{\epsilon^n},
\end{align}
where $\mu$ is an arbitrary scale.
From the $\mu$ independence of the bare parameters, one can define the $\beta$-functions of each parameter as
\begin{align}
m_a^2\beta_{m_a^2} &= \lim_{\epsilon\to0} \mu\frac{d m_a^2}{d\mu} = \sum_{k,b}b_{ab,k}^{(1)}g_km_b^2, \\
 \beta_\Omega & =\lim_{\epsilon\to0} \mu\frac{d \Omega}{d\mu} = \omega_1,
\end{align}
where $b_{ab,k}^{(1)}=db_{ab}^{(1)}/dg_k$. It is important to note that the $\beta$-functions are given by the coefficients of the single $\epsilon$ pole, which implies that if those coefficients are modified by thermal resummations, the $\beta$-functions would no longer remain the same for the theoretical consistency. This is exactly the case we consider in the following.   

In resummed perturbation theories, the Lagrangian is reorganized as~\cite{Parwani:1991gq}
\begin{align}
\mathcal{L}_B &= \mathcal{L}_R+\mathcal{L}_{\text{CT}} \nonumber \\
&= \left[\mathcal{L}_R-\frac{1}{2}\Sigma_a(T)\phi_a^2\right] +\left[\mathcal{L}_{\text{CT}} +\frac{1}{2}\Sigma_a(T)\phi_a^2\right],
\label{L_parwani}
\end{align}
where $\Sigma_a(T)$ denotes the thermal mass of the scalar $\phi_a$. 
At the leading order, $\Sigma_a(T)=\mathcal{O}(g_i T^2)$ with $g_i$ representing scalar quartic couplings.
Even though nothing has changed in the bare Lagrangian, $\Sigma_a(T)$ in the first square brackets is regarded as the zeroth order in the resummed perturbation theory while that in the second ones is part of the counterterm (CT) which is one order higher in this perturbative expansion (referred to as \textit{thermal counterterm} hereafter).  Because of this reorganization, the propagators of the scalars are temperature dependent, and one encounters temperature-dependent divergences when computing effective potentials at loop levels. Although such divergences must be canceled in the all-order calculation, they inevitably appear at a fixed order in the resummed perturbation theory. With this consideration, we modify Eq.~(\ref{mB}) as
\begin{align}
m_{Ba}^2 & = \left(\delta_{ab}+\sum_{n=1}^\infty\frac{b_{ab}^{(n)}(g)}{\epsilon^n} \right)m_b^2+\sum_{n=1}^\infty\frac{\tilde{b}_{ab}^{(n)}(g)}{\epsilon^n}\Sigma_b(T).
\end{align}
The order-by-order RG invariance is another concern after the thermal resummation.
There may exist resummed perturbation theories in which $\Sigma_a(T)$ is self-consistently determined~\cite{Kneur:2015uha,*Kneur:2015moa}.
In our approach, however, $\Sigma_a(T)$ is predetermined as a solution to the gap equation, and not by the perturbation theory considered here~\cite{Funakubo:2023eic}. For a practical purpose, we use $\Sigma_a(T)$ that is shown to be scale invariant up to the two-loop level in the $\overline{\text{MS}}$ scheme (for details, see Appendix B in Ref.~\cite{Funakubo:2023eic}). The residual scale dependence of $\Sigma_a(T)$ is a matter of precision in its computation and can be improved by including higher-order terms in the gap equation. From our perspective, this is a separate matter from the scale dependence issue of the effective potential that we will discuss below, and we do not combine the two different scale dependencies for consistency. By virtue of the scale invariance of $\Sigma_a(T)$, we can choose the couplings at a particular fixed scale $g_i(\mu_\text{fixed})$ for $\Sigma_a(T)$. For simplicity, we employ initial values of the RG running for $\Sigma_a(T)$ and a high-temperature approximation, which is explained in detail in Appendix~B of Ref.~\cite{Funakubo:2023eic}.
We refer to $d\Sigma_a(T)/d\mu=0$ as the \textit{consistency condition} and prove the order-by-order RG invariance of the resummed effective potentials up to the two-loop level.
Following the same procedure as in the $\overline{\text{MS}}$ scheme with the consistency condition, one obtains
\begin{align}
m_a^2\beta_{m_a^2} =\sum_{k,b}\big(b_{ab,k}^{(1)}m_b^2+\tilde{b}_{ab,k}^{(1)}\Sigma_b\big)\sigma_kg_k.\label{beta_m2}
\end{align}
We note that although the vacuum energy is also modified by the thermal resummation, the relation $\beta_\Omega=\omega_1$
still holds under the aforementioned consistency condition. 

%---------------------------------------------------------- Section 3 ----------------------------------------------------------
\section{$\phi^4$ theory} \label{sec:phi4}
%----------------------------------------------------------------------------------------------------------------------------------
We demonstrate how our RG scheme works using the $\phi^4$ theory. The bare Lagrangian is given by
\begin{align}
\mathcal{L}_B &= \frac{1}{2}\partial_\mu \Phi_B \partial^\mu \Phi_B-V_B(\Phi_B),\\
V_B(\Phi_B) &= \Omega_B-\frac{\nu_B^2}{2}\Phi^2+\frac{\lambda_B}{4!}\Phi_B^4.\label{bLag_phi4}
\end{align}
As mentioned in Sec.~\ref{sec:beta}, after decomposing $\mathcal{L}_B$ into $\mathcal{L}_R$ and $\mathcal{L}_{\text{CT}}$, we subtract and add $\Sigma(T)$ in each part. The explicit forms of CTs are summarized in Ref.~\cite{Funakubo:2023eic}.
With the resummed Lagrangian, we evaluate the effective potential up to the two-loop level. 
Denoting the classical background field as $\varphi$, the tree-level effective potential has the form
\begin{align}
V_0(\varphi)&=\Omega+\frac{1}{2}\left(-\nu^2+\Sigma(T)\right)\varphi^2+\frac{\lambda\mu^\epsilon}{4!}\varphi^4.
\end{align}
The field-dependent mass is defined as
\begin{align}
M^2 = \frac{\partial^2 V_0}{\partial \varphi^2}=m^2 + \Sigma(T), 
 \end{align}
with $m^2 = -\nu^2+\lambda\mu^\epsilon \varphi^2/2$. Using a propagator with $M^2$, one can obtain the one-loop correction to the effective potential~\cite{Parwani:1991gq}
\begin{align}
\mu^\epsilon V_1(\varphi)&= \frac{M^4}{4(16\pi^2)}\left(-\frac{2}{\epsilon}+\ln\frac{M^2}{\bar{\mu}^2}-\frac{3}{2}+\mathcal{O}(\epsilon)\right),\label{V1phi4}
\end{align}
where $\bar{\mu} = \sqrt{4\pi e^{-\gamma_E^{}}} \mu \simeq 2.66 \mu$ with $\gamma_E^{}$ being the Euler constant. In our renormalization scheme, we remove the divergences including the temperature-dependent pieces by the one-loop CTs, resulting in  
\begin{align}
\delta^{(1)}\Omega &= \frac{1}{\epsilon}\frac{(\nu^2-\Sigma)^2}{32\pi^2},\quad
\delta^{(1)}\nu^2 = \frac{1}{\epsilon}\frac{\lambda(\nu^2-\Sigma)}{16\pi^2}, \nonumber\\
\delta^{(1)}\lambda &= \frac{1}{\epsilon}\frac{3\lambda^2}{16\pi^2}.
\end{align}
The bare $\nu_B^2$ is expressed as
\begin{align}
\nu_B^2 &= Z_\Phi^{-1}(\nu^2+\delta^{(1)} \nu^2),
\label{nu2B_1L}
\end{align}
where $Z_\Phi$ denotes the wavefunction renormalization constant of $\Phi$, and $Z_\Phi=1$ at the one-loop level. From Eq.~(\ref{nu2B_1L}), the coefficient of the single $\epsilon$ pole is found to be $b_1(\lambda)=-\tilde{b}_1(\lambda)=\lambda/16\pi^2$.  Plugging them into the formula (\ref{beta_m2}), one obtains
\begin{align}
\nu^2\beta_{\nu^2}^{(1)} = \frac{\lambda(\nu^2-\Sigma)}{16\pi^2}.\label{beta1_nu2}
\end{align}
By doing the same step, one can find the $\beta$-functions of the remaining parameters and $\gamma$-function as
\begin{align}
\beta_\Omega^{(1)} & = \frac{(\nu^2-\Sigma)^2}{32\pi^2},\quad
\beta_\lambda^{(1)} =  \frac{3\lambda^2}{16\pi^2},\quad \gamma_\Phi^{(1)} = 0.
\end{align}
Note that the $\beta$-functions in our scheme are reduced to those in the $\overline{\text{MS}}$ scheme by taking $\Sigma=0$, 
which implies that differences between our scheme and $\overline{\text{MS}}$ scheme could be sizable when $\Sigma$ dominates over $\nu^2$.
We also note that in the $\overline{\text{MS}}$ scheme, the temperature-dependent divergences appearing in Eq.~(\ref{V1phi4}) remain at this order, and higher-order loop contributions are needed to cancel them~\cite{Karsch:1997gj} (See also Ref.~\cite{Laine:2017hdk}).\footnote{If $\mathcal{L}_R$ and $\mathcal{L}_{\text{CT}}$ are defined as in Ref.~\cite{Banerjee:1991fu,Chiku:1998kd} instead of the way they are defined in Eq.~(\ref{L_parwani}), the order-by-order renormalization with the $\overline{\text{MS}}$ scheme also holds by regarding the thermal mass term as one-order higher.
}

After the renormalization, the resummed one-loop effective potential is given by
\begin{align}
V_{\text{eff}}(\varphi) = V_0(\varphi)+V_1(\varphi),
\end{align}
where
\begin{align}
V_0(\varphi)&=\Omega+\frac{1}{2}\left(-\nu^2+\Sigma(T)\right)\varphi^2+\frac{\lambda}{4!}\varphi^4, \label{RV0} \\
V_1(\varphi) &= \frac{M^4}{4(16\pi^2)}
\left(
	\ln\frac{M^2}{\bar{\mu}^2}-\frac{3}{2}
\right)
+\frac{T^4}{2\pi^2}I_B(A^2)-\frac{1}{2}\Sigma(T)\varphi^2,\label{RV1}
\end{align}
with $A^2=M^2/T^2$ and the thermal function of the boson ($I_B$) is defined as
\begin{align}
I_B(A^2) = \int_0^\infty dx~x^2\ln\left(1-e^{-\sqrt{x^2+A^2}}\right).
\end{align}
The last term in $V_1(\varphi; T)$ is nothing but the thermal CT. In the high-$T$ expansion, the $+\Sigma(T)\varphi^2/2$ term arises from $T^4I_B/(2\pi^2)$, which is canceled by the thermal CT, avoiding the double counting of $\Sigma(T)\varphi^2/2$.  

As is the one-loop level, we regularize the two-loop effective potential by requiring that all the divergences be absorbed by the CTs, As a result, the two-loop contributions to the $\beta$-functions of the model parameters in our scheme are, respectively, given by
\begin{align}
\gamma_\Phi^{(2)} &= \frac{\lambda^2}{12(16\pi^2)^2}, \\
\beta_\Omega^{(2)} & = \frac{(\nu^2-\Sigma)\Sigma}{16\pi^2}, \\
\nu^2\beta^{(2)}_{\nu^2} &= \frac{\lambda^2(-\nu^2+\Sigma)}{(16\pi^2)^2}+\frac{\lambda\Sigma}{16\pi^2}+2\nu^2\gamma_\Phi^{(2)},
\label{beta2_nu2} \\
\beta_\lambda^{(2)} &=  -\frac{6\lambda^3}{(16\pi^2)^2}+4\lambda\gamma_\Phi^{(2)}.
\end{align}
We should note that $\beta_{\nu^2}^{(2)}$ contains the $\lambda\Sigma/(16\pi^2\nu^2)$ term that is only one-loop suppressed. This is exactly the same form as the thermal correction term in Eq.~(\ref{beta1_nu2}) with an opposite sign. Therefore, they are seemingly canceled with each other in the sum of the one- and two-loop $\beta$-functions $\beta_{\nu^2} = \beta_{\nu^2}^{(1)}+\beta_{\nu^2}^{(2)}$. However, when we evaluate $\beta_{\nu^2}$ perturbatively, $\lambda\Sigma/(16\pi^2\nu^2)$ in $\beta_{\nu^2}^{(2)}$ should be treated as the one-order higher term than that in $\beta_{\nu^2}^{(1)}$. 
In contrast to the $\overline{\text{MS}}$ scheme, $\beta_\Omega^{(2)}$ is nonzero in our scheme. 

After the renormalization, the two-loop correction to the resummed effective potential is cast into the form
\begin{align}
V_2(\varphi) &= 
\frac{\lambda}{8}\bar{I}^2(M)-\frac{\lambda^2\varphi^2}{12}\tilde{H}(M)-\frac{1}{2}\Sigma(T)\bar{I}(M),
\label{RV2}
\end{align}
where the thermal functions $\tilde{H}(M)$ and $\bar{I}(M)$ are defined in Ref.~\cite{Funakubo:2023eic}.
The last term comes from the thermal CT which plays a role in eliminating the double counting and linear-like terms in $\varphi$ such as $\mathcal{O}((M^2)^{1/2}T^3)$~\cite{Arnold:1992rz}.

Now we scrutinize the RG invariances of the resummed effective potentials obtained above. 
The effective potential satisfies RGE
\begin{align}
0 &= \mu\frac{ dV_{\text{eff}}}{d\mu} \equiv \mathcal{D}V_{\text{eff}} \nonumber\\
&= 
\left[
\mu\frac{\partial}{\partial \mu}+\nu^2\beta_{\nu^2}\frac{\partial}{\partial \nu^2}+\beta_\lambda\frac{\partial}{\partial\lambda}-\gamma_\Phi^{}\varphi\frac{\partial}{\partial \varphi}+\beta_\Omega\frac{\partial}{\partial\Omega}
\right]V_{\text{eff}}. \label{RGinv}
\end{align}

Let us check the RG invariance of the resummed effective potential at the one-loop level. Applying (\ref{RGinv}) to $V_0$ and $V_1$ respectively, one finds
\begin{align}
\mathcal{D}V_0|_{\text{one-loop}} & = \beta^{(1)}_\Omega-\frac{\nu^2}{2}\beta_{\nu^2}^{(1)}\varphi^2+\frac{\beta_\lambda^{(1)}}{4!} \varphi^4 
=\frac{M^4}{32\pi^2},\\
\mathcal{D}V_1|_{\text{one-loop}} &= \mu \frac{\partial V_1}{\partial \mu}=-\frac{M^4}{32\pi^2}+\mathcal{O}\left(\frac{1}{(16\pi^2)^2}\right),
\label{DV1T0}
\end{align}
where the consistency condition $\mathcal{D}\Sigma=0$ is used. 
Therefore, one gets $\mathcal{D}(V_0+V_1)=0+\mathcal{O}(1/(16\pi^2)^2)$, and the error is the two-loop order. 
On the other hand, if one uses the $\overline{\text{MS}}$ scheme, the error is estimated as
$\mathcal{D}(V_0+V_1)_{\overline{\text{MS}}}=(-2m^2+\Sigma)\Sigma/(32\pi^2)+\mathcal{O}(1/(16\pi^2)^2) \to -\lambda\varphi^2\Sigma/(32\pi^2)+\mathcal{O}(1/(16\pi^2)^2)$, where the $\varphi$-independent terms are suppressed after the right arrow. Note that despite the lack of the RG invariance in the $\overline{\text{MS}}$ scheme, the scale dependence could be unexpectedly smaller than that in our scheme due to an accidental cancellation between the two different errors. An example is given in Ref.~\cite{Funakubo:2023eic}. However, such a less scale dependence has no robust footing.

The proof of the RG invariance at the two-loop level is also straightforward. Applying the derivative operator $\mathcal{D}$ to the resummed effective potentials (\ref{RV0}), (\ref{RV1}), and (\ref{RV2}), respectively, we can verify that $\mathcal{D}(V_0+V_1+V_2)|_{\text{two-loop}} =0+\mathcal{O}(1/(16\pi^2)^3)$. We here emphasize again that the order-by-order RG invariance holds by virtue of the modified $\beta$-functions in our scheme. 

Now we consider a further refinement that fully exploits the RG invariance to incorporate a series of temperature-dependent higher-order terms. The explicit form of the resummed one-loop effective potential that satisfies RGE (\ref{RGinv}) is 
\begin{align}
\lefteqn{\bar{V}_{\text{eff}}(\bar{\varphi};t)=\bar{V}_0(\bar{\varphi};t)+\bar{V}_1(\bar{\varphi};t) }\nonumber\\
 & = \bar{\Omega}+\frac{1}{2}\left(-\bar{\nu}^2+\Sigma\right)\bar{\varphi}^2+\frac{\bar{\lambda}}{4!}\bar{\varphi}^4 \nonumber\\
& \quad+\frac{\bar{M}^4}{4(16\pi^2)}\left(\ln\frac{\bar{M}^2}{e^{2t}\bar{\mu}_0^2}-\frac{3}{2}\right)+\frac{T^4}{2\pi^2}I_B(\bar{A}^2) -\frac{1}{2}\Sigma\bar{\varphi}^2,
\label{RGVeff}
\end{align}
with $\bar{A}=\bar{M}/T$, and $\bar{M}^2 = -\bar{\nu}^2+\Sigma(T)+\bar{\lambda}\bar{\varphi}^2/2$.
The barred parameters $\bar{\Omega}$, $\bar{\nu}^2$, $\bar{\lambda}$, and $\bar{\varphi}$ are the running parameters as functions of $t=\ln(\bar{\mu}/\bar{\mu}_0)$ with $\bar{\mu}_0$ being an initial scale.
Hereafter, the unbarred parameters are defined at $t=0$. Because $t$ is arbitrary, it would be preferable to determine it in such a way that dominant higher-order terms are incorporated into the potential (\ref{RGVeff}). At zero temperature, we could choose $t(\varphi)=\ln(\bar{m}^2/\bar{\mu}_0^2)/2$ to absorb logarithmic terms that could ruin the validity of perturbativity in some domain~\cite{Kastening:1991gv, Bando:1992np,*Bando:1992wy}. At finite temperature, however, this choice is not able to tame dominant temperature-dependent terms arising from
\begin{align}
\bar{I}(\bar{M}) \mathop{\simeq}_{T\gg \bar{M}} \frac{T^2}{12},
\end{align}
For this reason and because the truncation error of RGE at this order is given by
\begin{align}
\frac{d \bar{V}_{\text{eff}}(\bar{\varphi};t)}{d t}=\frac{\partial \bar{V}_{\text{eff}}(\bar{\varphi};t)}{\partial t}
& = 0+\frac{1}{2}\frac{\partial\bar{M}^2}{\partial t}\bar{I}(\bar{M}),
\end{align}
we choose $t$ to eliminate this error at each $\varphi$, yielding
\begin{align}
t(\varphi) &= \frac{8\pi^2}{\bar{M}^2}\bar{I}(\bar{M})_{t=0}.
\label{t-phi}
\end{align}
In this scheme, the higher-order terms in $\bar{I}(\bar{M})$ appearing beyond the one-loop order can be taken into (\ref{RGVeff}) through the $t$-$\varphi$ relation in Eq.~(\ref{t-phi}).
In the zero temperature limit, Eq.~(\ref{t-phi}) is reduced to $t(\varphi)=\ln(\bar{m}^2/e\bar{\mu}_0^2)/2$. Therefore, our scheme in this limit is related to the aforementioned scheme $t(\varphi)=\ln(\bar{m}^2/\bar{\mu}_0^2)/2$ by changing the input scale $\bar{\mu}_0$ to $\bar{\mu}_0/\sqrt{e}$.

Let us denote the RG-improved potential (\ref{RGVeff}) with $\ell$-loop order $\beta$-functions as $\bar{V}_{\text{eff}}^{(\ell)}(\varphi; t(\varphi))$, which contains some of the higher-order terms beyond the $\ell$-loop, arising from the running parameters even including the vacuum energy $\bar{\Omega}(t)$.
It is easy to check that $\bar{V}_{\text{eff}}^{(1)}(\varphi; t(\varphi))$ include $\bar{I}(\bar{M})$ terms in the two-loop effective potential (\ref{RV2}) using the $t$ expansion of (\ref{RGVeff})
\begin{align}
\bar{V}_{\text{eff}}(\bar{\varphi}; t) &= \bar{V}_{\text{eff}}(\varphi; 0)
+\frac{\partial \bar{V}_{\text{eff}}(\bar{\varphi}; t)}{\partial t}\bigg|_{t=0}t \nonumber\\
&\quad +\frac{1}{2}\frac{\partial^2 \bar{V}_{\text{eff}}(\bar{\varphi}; t)}{\partial t^2}\bigg|_{t=0}t^2+\cdots
\label{appRGVeff}
\end{align}
and the $t$-$\varphi$ relation (\ref{t-phi}). 
From those expressions, for example, it follows that 
\begin{align}
\bar{V}_{\text{eff}}^{(1)}(\bar{\varphi}; t(\varphi))
&=\bar{V}_{\text{eff}}^{(1)}(\varphi; 0)
+\frac{\lambda(M^2+\lambda\varphi^2)}{8M^2}\bar{I}^2(M)_{t=0}.
\end{align}
The second term is exactly the same as $\mathcal{O}(\bar{I}^2(M))$ terms in $V_2(\varphi)$ given in Eq.~(\ref{RV2}).
On the other hand, $\bar{V}_{\text{eff}}^{(2)}(\bar{\varphi}; t(\varphi))$ contains even $\mathcal{O}(\bar{I}(M))$ terms including the thermal CT in $V_2(\varphi)$. This appears analogous to the leading and next-to-leading logarithmic resummations at zero temperature~\cite{Kastening:1991gv, Bando:1992np,*Bando:1992wy}. An important difference is that $t$-expanded $\bar{V}_{\text{eff}}^{(2)}(\bar{\varphi}; t(\varphi))$ includes more terms that are not present in the fixed-order ($t=0$) $V_2(\varphi)$ in Eq.~(\ref{RV2})~\cite{Funakubo:2023eic}.
Since the $\phi^4$ theory does not accommodate the first-order phase transition, we will consider a multi-scalar theory in the next section.
\section{$\phi^4$ theory with additional scalar}
As the simplest extension, another real scalar field is added to the $\phi^4$ theory in order to compare quantities related to first-order phase transition in both $\overline{\text{MS}}$ and our schemes.
For illustration, we consider a simplified potential by imposing two $\mathbb{Z}_2$ symmetries.
The bare potential of the extended model has the form
\begin{align}
V_0(\Phi_{B1},\Phi_{B2})&=\Omega_B+\frac{\nu_{B1}^2}{2}\Phi_1^2+\frac{\nu_{B2}^2}{2}\Phi_{B2}^2 \nonumber\\
&\quad+\frac{\lambda_{B1}}{4!}\Phi_{B1}^4+\frac{\lambda_{B2}}{4!}\Phi_{B2}^4+\frac{\lambda_{B3}}{4}\Phi_{B1}^2\Phi_{B2}^2,
\end{align}
which is invariant under $\mathbb{Z}_2$ symmetries $\Phi_{B1}\to -\Phi_{B1}$ and $\Phi_{B2}\to-\Phi_{B2}$.
As in the $\phi^4$ theory, we subtract and add the thermal masses of $\Phi_1$ and $\Phi_2$ (denoted as $\Sigma_1$ and $\Sigma_2$) in the renormalized Lagrangian and CTs, respectively. In this study, we assume that only $\Phi_1$ develops the vacuum expectation value while $\Phi_2$ does not. For later use, the classical background field of $\Phi_1$ is denoted as $\varphi$.
It is straightforward to show that the finite temperature effective potentials up to the two-loop level satisfy RGE by virtue of the temperature-dependent $\beta$-functions in our scheme~\cite{Funakubo:2023eic}. 
To improve the potentials further, we choose $t$ in order to incorporate a series of temperature-dependent higher-order terms.
For instance, at the one-loop order, we impose 
\begin{align}
\frac{\partial \bar{V}_{\text{eff}}(\bar{\varphi}; t)}{\partial t}
= 0+\frac{1}{2}\sum_i\frac{\partial \bar{M}_i^2}{\partial t}\bar{I}(\bar{M}_i) =0,
\end{align}
where $\bar{M}_1^2 = \bar{\nu}_1^2+\Sigma_1(T)+\bar{\lambda}_1\bar{\varphi}^2/2$ and $\bar{M}_2^2 = \bar{\nu}_2^2+\Sigma_2(T)+\bar{\lambda}_3\bar{\varphi}^2/2$ with $\Sigma_1(T) = (\lambda_1+\lambda_3)T^2/24$ and $\Sigma_2(T) = (\lambda_2+\lambda_3)T^2/24$. With this condition, the RG-improved effective potential is given by
\begin{align}
\lefteqn{\bar{V}_{\text{eff}}(\bar{\varphi};t(\varphi)) = \bar{V}_0(\bar{\varphi};t(\varphi))+\bar{V}_1(\bar{\varphi};t(\varphi))} \nonumber\\
& = \bar{\Omega}+\frac{1}{2}\left(\bar{\nu}_1^2+\Sigma_1(T)\right)\bar{\varphi}^2+\frac{\bar{\lambda}_1}{4!}\bar{\varphi}^4 \nonumber \\
& \quad +\sum_{i=1,2}\left[\frac{\bar{M}_i^4}{4(16\pi^2)}\left(\ln\frac{\bar{M}_i^2}{e^{2t}\bar{\mu}_0^2}-\frac{3}{2}\right)+\frac{T^4}{2\pi^2}I_B(\bar{A}_i^2)\right] \nonumber\\
&\quad -\frac{1}{2}\Sigma_1(T)\bar{\varphi}^2,
\label{RGVeff_exphi4}
\end{align}
where $\bar{A}_i=\bar{M}_i/T$, and the explicit form of $t(\varphi)$ is  
\begin{align}
t(\varphi) & =\frac{ 8\pi^2\sum_i\frac{\partial\bar{M}_i^2}{\partial t}\bar{I}(\bar{M}_i)_{t=0} }{\sum_i\bar{M}_i^2\frac{\partial\bar{M}_i^2}{\partial t}}.
\label{t-phi_exphi4}
\end{align}
Expanding (\ref{RGVeff_exphi4}) in powers of $t$, $\bar{V}_{\text{eff}}^{(1)}(\bar{\varphi}; t)$ is cast into the form
\begin{align}
\bar{V}_{\text{eff}}^{(1)}(\bar{\varphi}; t(\varphi))
=\bar{V}_{\text{eff}}^{(1)}(\varphi;0)
+\frac{ \big( \sum_i\alpha_i\bar{I}(M_i)_{t=0} \big)^2 }{8\sum_i\alpha_iM_i^2},
\label{RGVeff1}
\end{align}
where $\alpha_i = 16\pi^2\partial \bar{M}_i^2/\partial t|_{t=0}$. Unlike the $\phi^4$ theory, the form of the second term does not coincide with that in the fixed-order two-loop effective potential $V_2$. Such a mismatch between the RG-improved and fixed-order effective potentials is peculiar to the multi-field case, which is attributed to the fact that the single parameter $t$ alone cannot incorporate two different $\bar{I}(M_i)$ terms correctly in principle. We investigate to what extent our scheme can capture the higher-order effects by comparing with the two-loop order result.
%------------------------------------------------------------------------------------------------------
\begin{figure*}[t]
\begin{center}
\includegraphics[width=7cm]{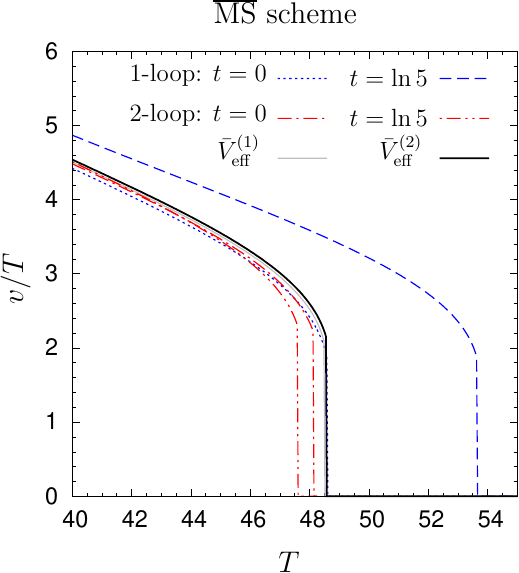}\hspace{1cm}
\includegraphics[width=7cm]{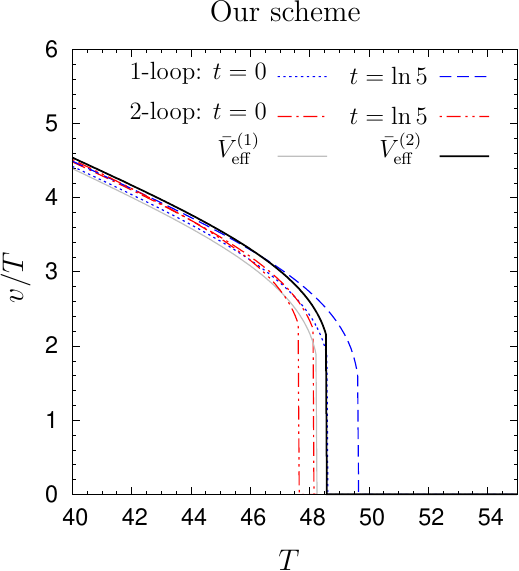} 
\caption{$v(T)/T$ as a function of $T$ in the $\overline{\text{MS}}$ scheme (Left) and our scheme (Right). We take $v({\bar{\mu}_0})=200$, $m_{\phi_1}({\bar{\mu}_0})=5.0$, $m_{\phi_2}(\bar{\mu}_0)=125$, $\nu_2^2(\bar{\mu}_0)=85.0^2$,  $\lambda_2(\bar{\mu}_0)=5.0$, where $\nu_1^2(\bar{\mu}_0)$ and $\lambda_1(\bar{\mu}_0)$ are determined by the first and second derivatives of the effective potentials at a given order while $\lambda_3(\bar{\mu}_0)$ at the tree-level. $\bar{\mu}_0$ is fixed by the condition $t(\varphi=v)=0$. At the both one- and two-loop levels, $\bar{\mu}_0\simeq 75.81$. The dimensionful parameters are expressed in units of any mass scale.
}
\label{fig:vcTc_T}
\end{center}
\end{figure*}
%------------------------------------------------------------------------------------------------------

In this model, there are 5 parameter in the scalar potential, i.e., $(\nu_1^2, \nu_2^2, \lambda_1, \lambda_2, \lambda_3)$. Using vacuum and mass conditions, we convert them into $(v, \nu_2^2, m_{\phi_1}, \lambda_2, m_{\phi_2})$.
As an example of the first-order phase transition, we take $v({\bar{\mu}_0})=200$, $m_{\phi_1}({\bar{\mu}_0})=5.0$, $m_{\phi_2}(\bar{\mu}_0)=125$, $\nu_2^2(\bar{\mu}_0)=85.0^2$,  $\lambda_2(\bar{\mu}_0)=5.0$, where $\nu_1^2(\bar{\mu}_0)$ and $\lambda_1(\bar{\mu}_0)$ are determined by the first and second derivatives of the effective potentials at a given order while $\lambda_3(\bar{\mu}_0)$ at the tree level. $\bar{\mu}_0$ is fixed by the condition $t(\varphi=v)=0$. At the both one- and two-loop levels, $\bar{\mu}_0\simeq 75.81$. The dimensionful parameters are given in units of any mass scale.
 Because of the smallness of $m_{\phi_1}$, the appearance of the imaginary parts of the effective potentials is only limited to low temperature, and the effective potentials are all real and well-defined near critical temperatures $T_C$, where the potentials have two degenerate minima.

In Fig.~\ref{fig:vcTc_T}, $v(T)/T$ are shown as a function of the temperature $T$ in the $\overline{\text{MS}}$ (left) and our (right) schemes, respectively. 
The dotted and dashed curves in blue represent the results obtained by using the one-loop effective potential (\ref{RGVeff_exphi4}) in the cases of $t=0$ and $\ln 5$, respectively. The intersections between the horizontal axis and each curve represent $T_C$.
As clearly seen, the renormalization scale dependence on $T_C$ in the $\overline{\text{MS}}$ case is much larger than that in our scheme. This is due to a large violation of RG invariance in the former. On the other hand, the dot-dashed and two-dot-dashed lines in red correspond to the results using (\ref{RGVeff_exphi4}) and the two-loop effective potential $\bar{V}_2(\bar{\varphi},t)$ [the RG-improved version of Eq.~(\ref{RV2}) but with the two scalars] with $t=0$ and $\ln 5$, respectively. In those cases, the renormalization scale dependence is even milder than that in the one-loop result with our scheme. Note that the improvement in the $\overline{\text{MS}}$ scheme is because of the \textit{partial} restoration of the RG invariance. One can explicitly check that the effective potential follows the RG invariance up to the $\mathcal{O}(\lambda_i^2T^2)$ order in the high-temperature limit~\cite{Arnold:1992rz}. In this parameter set, the residual RG-noninvariant terms are numerically small and the truncation errors become dominant, which explains the two-loop results.
We also overlay the results by use of the effective potentials $\bar{V}_{\text{eff}}^{(1)}(\bar{\varphi}, t(\varphi))$ and $\bar{V}_{\text{eff}}^{(2)}(\bar{\varphi}, t(\varphi))$ with the $t$-$\varphi$ relation (\ref{t-phi_exphi4}). The former is denoted by the solid line in grey while the latter by thick solid line in black. One can see that $v(T_C)/T_C$ using $\bar{V}_{\text{eff}}^{(2)}(\bar{\varphi}, t(\varphi))$ in both schemes yield $v(T_C)/T_C\simeq 2.2$, which lies within the two-loop level scale uncertainties, i.e., $2.1\lesssim v(T_C)/T_C\lesssim 2.3$ ($\overline{\text{MS}}$ scheme) and $2.2\lesssim v(T_C)/T_C\lesssim 2.3$ (our scheme), while in the cases using $\bar{V}_{\text{eff}}^{(1)}(\bar{\varphi}, t(\varphi))$, it is found that $v(T_C)/T_C\simeq 1.9~(2.1)$ in our ($\overline{\text{MS}}$) scheme, respectively.
This demonstration suggests that $\bar{V}_{\text{eff}}^{(2)}(\bar{\varphi}, t(\varphi))$ can give the results closer to those at the two-loop order.  

%---------------------------------------------------------- Section 5 ----------------------------------------------------------
\section{Conclusion}
%----------------------------------------------------------------------------------------------------------------------------------
We have proposed the novel method for renormalization group improvement of thermally resummed effective potential.
In our method, the RG invariance of the resummed finite-temperature effective potential holds order by order since the $\beta$-functions are correctly defined in resummed perturbation theory. Taking the extended $\phi^4$ theory as an example, we showed that the renormalization scale dependence of the first-order phase transition quantities, especially $T_C$ in our scheme is much smaller than that in the $\overline{\text{MS}}$ scheme even at the one-loop level. At the two-loop level, however, no significant differences are observed in both schemes. This is because that the RG invariance in the $\overline{\text{MS}}$ is restored up to $\mathcal{O}(\lambda_i^2T^2)$ order in the high temperature limit and the residual RG-noninvariant terms are numerically unimportant. 
We also devised the tractable method that enables one to incorporate a series of temperature-dependent higher-order corrections utilizing the RG invariance in our scheme. Applying this method to RG-improved one-loop effective potential, $v(T_C)/T_C$ in the case of $\bar{V}_{\text{eff}}^{(2)}(\bar{\varphi}, t(\varphi))$ falls within the errors of the two-loop order renormalization scale dependence, suggesting that our refined method could be a practical choice when the two-loop effective potential is not available.

%\begin{acknowledgments}
%\end{acknowledgments}

%\appendi
%-------------------------------------------------------------------------------------------------------------------------------
%
\bibliography{refs}
%
%-------------------------------------------------------------------------------------------------------------------------------

\end{document}